# BEAM SCREEN ISSUES

E. Métral, CERN, Geneva, Switzerland

*Abstract*

In the High Energy LHC (HE-LHC), a beam energy of about 16.5 TeV is currently contemplated. The beam screen issues linked to the use of 20 T dipole magnets instead of 8.33 T are discussed, with a particular emphasis on two mechanisms, the magneto-resistance and the anomalous skin effect, assuming the nominal machine and beam parameters. The magneto-resistance effect always leads to an increase of the material resistivity (as the mean free path in the presence of a transverse magnetic field becomes smaller). As concerns the anomalous skin effect, the anomalous increase of surface resistance of metals at low temperatures and high frequencies is attributed to the long mean free path of the conduction electrons: when the skin depth becomes much smaller than the mean free path, only a fraction of the conduction electrons moving almost parallel to the metal surface is effective in carrying the current and the classical theory breaks down.

## INTRODUCTION

In the LHC, about 90% (i.e. the beam screen) is maintained between 5 and 20 K, while the other 10% is at room temperature (with a 2 mm thick copper beam pipe). The main purpose of the beam screen is to shield the cold bore from the synchrotron radiation and it is made of stainless steel to resist to the mechanical stresses. A copper coating with a thickness of 75 μm is used to keep the resistivity as low as possible for the transverse resistive-wall coupled-bunch instability [1]. The latter is a low-frequency phenomenon, from a few kHz to a few MHz, where the Magneto-Resistance (MR) effect is important and must be correctly taken into account. The power loss is a more involved issue due, in addition, to the short bunch length, the Anomalous Skin Effect (ASE) and the surface roughness (both important at high frequencies). A much smaller copper thickness could have been chosen (of the order of 1 μm) if only this effect had to be taken into account. The drawback from copper coating is the eddy currents, which are mainly concentrated in the copper layer in the cases of magnet's quenches. Therefore, for the quench force consideration, which deforms the beam screen horizontally, the smaller the copper coating thickness the better.

It is worth mentioning that the other impedance issues carefully studied in the past were the pumping slots (needed for the vacuum) and the longitudinal weld. Furthermore, I will not discuss here (as it will be discussed elsewhere) the important issue of Synchrotron Radiation (SR), even if in the HE-LHC the power would be increased by ~ 30 (from ~ 3.8 kW for one beam to ~ 120 kW: the scaling goes with the fourth power of the energy) and the critical photon energy by ~ 13 (from ~ 43 eV to ~ 574 eV: the scaling goes with the magnetic field times the square of the energy), keeping all the other parameters constant.

In this paper, the current LHC beam screen is reviewed in Section 1. The MR effect is discussed in Section 2, recalling first what was done in the past, which was an approximation of the approximated Kohler's rule. The exact and approximate Kohler's rules are then discussed in some detail. Finally, Section 3 is devoted to the ASE, first reviewing what was done in the past, i.e. using the approximate formula, and then studying the exact formula from Reuter & Sondheimer.

## CURRENT LHC BEAM SCREEN

Figure 1 shows a beam screen design as it was built and installed in the LHC. It is worth mentioning that in the dipoles, some baffles (i.e. shields of the pumping slots) were installed (see Fig. 1), to avoid a direct e⁻ path along the magnetic field lines to the cold bore (which would then add to the heat load). For the arc beam screens, the inner dimension between flats (i.e. between the two flat parts of the beam screen) is 36.8 mm and the inner dimension between radii (i.e. between the two circular parts of the beam screen) is 46.4 mm. The stainless steel thickness is 1.0 mm and the copper coating thickness is 75 μm. For the LSS (Long Straight Section) beam screens, the inner dimension between flats varies between 37.6 mm and 61.0 mm, and the inner dimension between radii varies between 47.2 mm and 70.7 mm. The stainless steel thickness is 0.6 mm and the copper coating thickness is still 75 μm. The lengths of the slots needed for the vacuum pumping are 6, 7, 8, 9 or 10 mm, i.e. 8 mm on average. The width of the slots is 1.5 mm in the arcs and 1.0 mm in the LSS. Finally, the total surface covered by the holes is ~ 4.0% in the arcs, while it varies between ~ 1.8% and ~ 2.6% in the LSS, depending on the screen diameter.

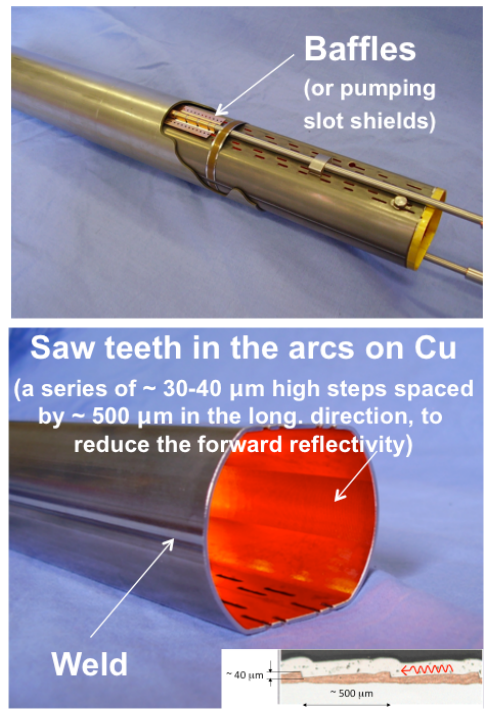

Figure 1: Beam screen as it was built and installed in the LHC (Courtesy of N. Kos).

The power loss from the induced currents in the beam screen (neglecting the holes) at 7 TeV/c is given by (the same numerical result is obtained with the more precise multi-layer impedance formula [2])

$$P_{loss/m}^{G,RW,1\text{layer}} = \frac{1}{2\pi R} \Gamma\left(\frac{3}{4}\right) \frac{M}{b} \left(\frac{N_b e}{2\pi}\right)^2 \sqrt{\frac{c \rho Z_0}{2}} \sigma_t^{-3/2} \quad (1)$$
$$\approx 85 \text{ mW/m},$$

where $R \approx 4243$ m is the average machine radius, $\Gamma(3/4) \approx 1.23$ the Euler gamma function, $M = 2808$ the number of bunches, $b = 18.4$ mm the beam screen half height, $N_b = 1.15\,10^{11}$ p/b the number of protons per bunch, $e$ the elementary charge, $c$ the speed of light, $\rho$ the resistivity (i.e. $5.5\,10^{-10}$ Ωm for copper at 20 K), $Z_0$ the free space impedance (i.e. 377 Ω), and $\sigma_t = 0.25$ ns is the bunch length. Note that the power loss goes with the square of the bunch charge, which means that it is ~ 2 times higher for the ultimate intensity ($1.7 \times 10^{11}$ p/b) than for the nominal one ($1.15 \times 10^{11}$ p/b).

The power loss from the induced currents in the weld are given by

$$P_{loss/m}^{Weld} \approx P_{loss/m}^{G,RW,1\text{layer}} \sqrt{\frac{\rho_{SS}^{20K}}{\rho_{Cu}^{20K}}} \frac{\Delta_l^{Weld}}{2\pi b} \approx 48 \text{ mW/m}, \quad (2)$$

with

$$\frac{\Delta_l^{Weld}}{2\pi b} = \frac{2}{2\pi\,18.4} = \frac{1}{\pi\,18.4} \approx \frac{1}{60}, \quad (3)$$

where $\rho_{SS}^{20K} = 6\times 10^{-7}$ Ωm. Therefore, even though the weld corresponds to only ~ 1/60 of the cross-section, the power loss due to the weld is not negligible at all and amounts to ~ 57% of the power loss without the weld.

If one compares the previous estimates with what was computed in the past for a single beam [3], we find that instead of the 85 mW/m a value of 110 mW/m (based on measurements of LHC dipole beam screen samples without magnetic field and subsequent extrapolation) is quoted (noting that the ASE, not yet taking into account here, gives an increase by ~ 11%). Note also that ~ 80 mW/m were obtained from simulations [4]. Concerning the weld, 10 mW/m were mentioned in Ref. [3] instead of the 48 mW/m computed in Eq. (2), while ~ 27 mW/m were found in Ref. [4]. Finally, ~ 1 mW/m is found for the most critical pumping holes in the arc beam screen (which is very close to the result of Ref. [5]), whereas 10 mW/m are mentioned in Ref. [3].

The transverse resistive-wall impedance in the classical regime, which is a good approximation in the present case, is given by

$$Z_\perp^{RW}(\omega) = (1+j) \frac{L Z_0}{\pi b^3} \sqrt{\frac{\rho}{2 \mu_0 \omega}}, \quad (4)$$

where $j$ is the imaginary unit, $L$ the longitudinal length of the structure, $\mu_0$ the vacuum permeability and $\omega$ is the angular frequency. It can be seen that it is proportional to the square root of the resistivity and that it goes with the inverse of the pipe radius to the power of three. The transverse impedance should be weighted by the transverse betatron function at the location of the impedance to correctly model the beam dynamics. Using the exact dimensions of all the beam screens and the correct local transverse betatron functions, the transverse coupled-bunch instabilities were studied and the results for the horizontal plane are shown in Fig. 2. It should be reminded that - Im (ΔQ) / $10^{-4}$ = 1 corresponds to a rise time of ~ 1600 turns, i.e. ~ 140 ms, and that the transverse feedback should be able to damp down to ~ 20–40 turns [6]. It can be seen from Fig. 2 that the beam screen contributes very little to the real part of the tune shift (which is dominated by all the collimators), but contributes significantly (~ 50%) to the imaginary part.

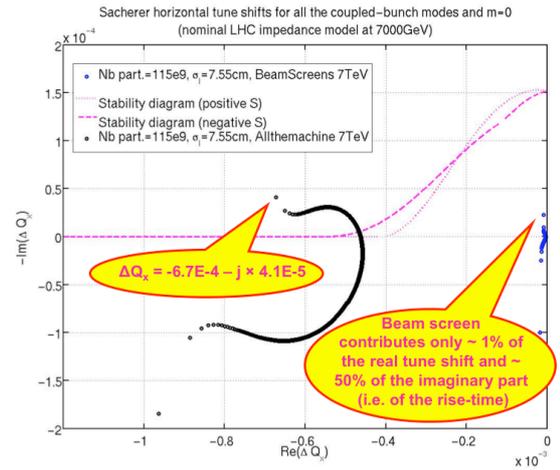

Figure 2: Horizontal tune shifts using Sacherer's formula [7] for all the coupled-bunch modes, for the head-tail mode 0 and for 0 chromaticity (using the LHC impedance model at 7 TeV/c). The two purple curves describe the stability diagrams with maximum otupoles' current [3]. Courtesy of N. Mounet.

# MAGNETO-RESISTANCE (MR)

How were the values of the beam screen copper resistivity at low (i.e. injection) and high (i.e. top energy) magnetic field obtained? In Ref. [8], the following formula, referred to as "Kohler's law", was used

$$\frac{\rho(B,T) - \rho_0(T)}{\rho_0(T)} = \frac{\Delta\rho}{\rho_0} = 10^{-2.69} (B\, RRR)^{1.055}, \quad (5)$$

where $B$ is the magnetic induction, $T$ the temperature, $\rho_0$ the resistivity at temperature $T$ but without magnetic field, and $RRR = R(273\,K)/R(T)$ is the Residual Resistance Ratio, which is a measure of purity of a material. Note that the resistance and resistivity are linked by the relation $R = \rho\, l / S$ (for long thin conductors), where $l$ is the length and $S$ the cross-sectional area, which means that $\Delta R / R_0 = \Delta\rho / \rho_0$.

As the resistivity decreases with temperature towards a minimum (determined by purity), the $RRR$ is sometimes defined as the ratio of the DC resistivity at room temperature to its cold-DC lower limit (see Fig. 3) [9].

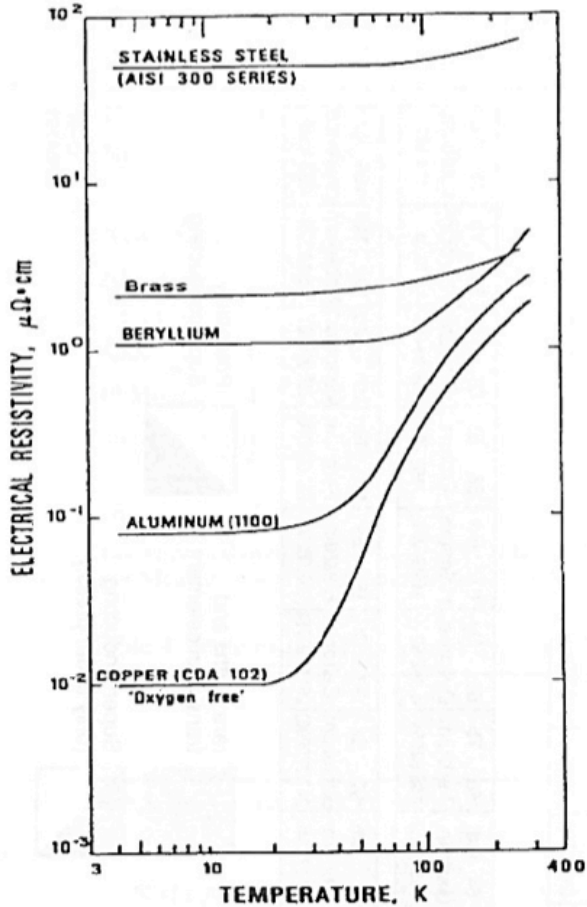

Figure 3: Resistivity of several metals vs temperature, in the absence of magnetic field.

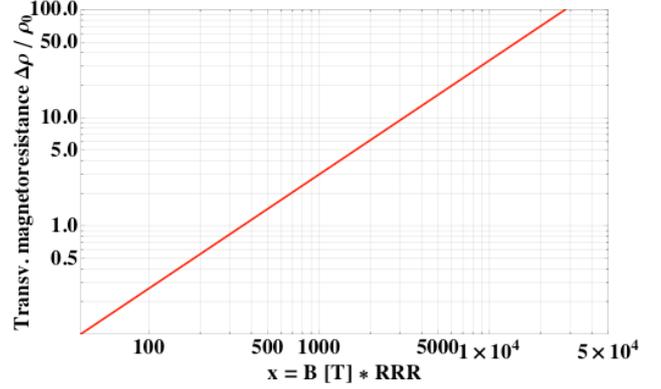

Figure 4: Plot of the approximate formula (of the approximate Kohler's rule) given by Eq. (5).

Assuming a $RRR$ of 100 and a resistivity at 20 K of $1.55\,10^{-10}$ Ωm, yields a resistivity at 20 K and 0.535 T (i.e. for the injection energy) of $1.8\,10^{-10}$ Ωm and a resistivity at 20 K and 8.33 T of $5.5\,10^{-10}$ Ωm. Using the same formula as Eq. (5) yields a resistivity at 20 K and 20 T of $11.2\,10^{-10}$ Ωm. The plot of the approximate formula (of the approximate Kohler's rule) is given in Fig. 4. For 0.535 T, $x = 53.5$ (see Fig. 4) and $\Delta\rho / \rho_0 \approx 0.14$. For 8.33 T, $x = 833$ and $\Delta\rho / \rho_0 \approx 2.5$. For 20 T, $x = 2000$ and $\Delta\rho / \rho_0 \approx 6.2$.

It is worth mentioning that in general care must be exercised when applying Kohler's rule to the magneto-resistance of some conductors (including high Tc-superconductors), where the density of charge carriers might change with temperature [10]. In fact, Kohler's rule may take two forms, one exact and one approximate. If there is only one relaxation rate in the transport process of a certain conductor, the exact Kohler's rule writes

$$\frac{\Delta\rho}{\rho_0} = F(H\,\tau), \quad (6)$$

which is generally a tensor, where $H = B/\mu_0$ is the magnetic field, $\tau$ the relaxation rate (or time) and $F$ is a function given only by the intrinsic electronic structure and external geometry of the conductor. The link between the relaxation time and the DC resistivity under 0 magnetic field can be found by using Ohm's law for a wire carrying a current density. The equation of motion for one electron is

$$m\frac{d\vec{v}}{dt} = -e\vec{E} - \alpha\vec{v}, \quad (7)$$

where $m$ is the electron mass, $\vec{v}$ the velocity, $t$ the time, $\vec{E}$ the electric field, and $\alpha = m/\tau$. In permanent (DC) regime, $d\vec{v}/dt = 0$ and $\vec{J} = -Ne\vec{v} = \sigma_{DC}\vec{E}$ is the current density, where $N$ is the density of carriers and

$$\rho_0 = \frac{1}{\sigma_{DC}} = \frac{m}{N e^2 \tau}. \tag{8}$$

The exact Kohler's rule of Eq. (6) can then be re-written

$$\frac{\Delta \rho}{\rho_0} = F\left(\frac{H}{\rho_0} \frac{m}{N e^2}\right). \tag{9}$$

If the factor $m/(N e^2)$ does not change with temperature, then Kohler's rule can be simplified to

$$\frac{\Delta \rho}{\rho_0} = F\left(\frac{B}{\rho_0}\right). \tag{10}$$

Equation (10) is Kohler's rule in its approximate but often used form. Most of the problem comes from $N$, which could be very sensitive to the temperature in various conductors. Equation (10) can be rewritten

$$\frac{\Delta \rho}{\rho_0} = F(B\ RRR), \tag{11}$$

as

$$\rho_0 = \rho_0(T) \propto \frac{1}{RRR}. \tag{12}$$

Equation (11) is the form of Kohler's law used for instance in Ref. [9], where the corresponding plot is shown in Fig. 5.

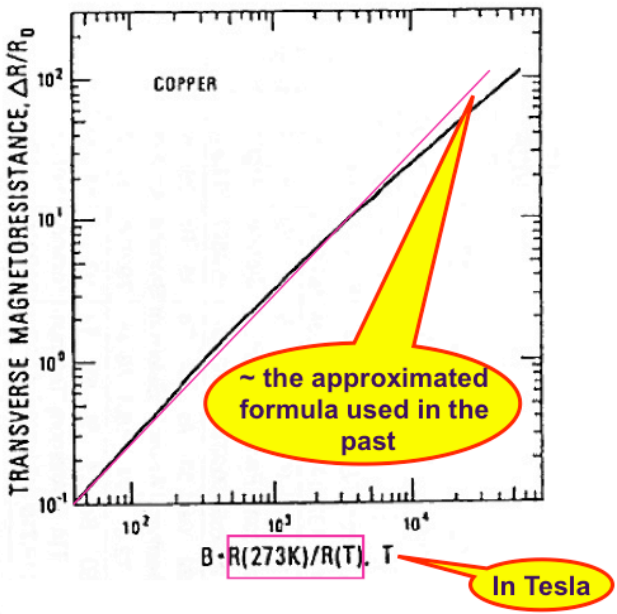

Figure 5: Kohler's plot for copper [9].

Experimental observations tell us that there is always an increase in resistance when the magnetic field is increased and that for small magnetic fields the resistance increase is proportional to the square of the magnetic field, whereas it becomes linear for very high magnetic fields. Note that Aluminum is an interesting material as concerns the magneto-resistance as it is one of the few materials which deviate from Kohler's rule [11] (a kind of saturation is observed; however, the secondary emission yield is very high, which prevents it from being used in machines where electron clouds could develop). But, why is there always an increase in resistance with increasing magnetic field? To answer this question it is useful to introduce two parameters, the mean free path of the electrons and the cyclotron radius. The mean free path $\lambda = \lambda(0)$ of a particle in the absence of magnetic field, is the average distance covered by a particle between successive impacts: $\lambda = \upsilon\ \tau$. This leads to

$$\lambda = \frac{m\ \upsilon}{e^2\ N\ \rho_0}. \tag{13}$$

As concerns the cyclotron radius, a particle, with a constant energy, describes a circle in equilibrium between the centripetal magnetic force and the centrifugal force, which leads to the cyclotron radius

$$r = \frac{m\ \upsilon}{e\ B}. \tag{14}$$

It can be seen from Eqs. (13) and (14), that

$$\frac{B}{\rho_0} \propto \frac{\lambda}{r}. \tag{15}$$

The case of a small transverse magnetic field is described in Fig. 6, where it can be seen that a smaller mean free path in the direction of motion is obtained, which means a larger resistivity (see Eq. (13)). Using the Taylor expansion of the sin function up to the second term, the mean free path can be approximated by

$$\lambda(H) \approx \lambda(0)\left\{1 - \frac{1}{6}\left[\frac{\lambda(0)}{r}\right]^2\right\}, \tag{16}$$

which leads to

$$\frac{\Delta \rho}{\rho_0} = -\frac{\Delta \lambda}{\lambda_0} \propto \left[\frac{\lambda(0)}{r}\right]^2 \propto \left[\frac{B}{\rho_0}\right]^2. \tag{17}$$

Equation (17) reveals indeed that for a small transverse magnetic field, the increase in resistivity due to the magneto-resistance is proportional to the square of the magnetic field.

Electrical measurements of beam screen wall samples in magnetic fields were performed in Ref. [13], which revealed that the trend line slopes of the voltage for all samples were always higher than the theoretical curves by

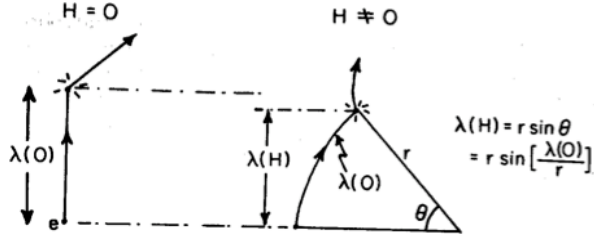

Figure 6: Reduction of the mean free path in the direction of motion in the presence of a transverse magnetic field. Courtesy of Jeff Fitzgerald [12].

~ 20% (see Fig. 7). These results confirmed the assumption of a heterogeneous *RRR* in the co-laminated copper layer: the copper close to the steel gets contaminated during the fabrication process such that the surface impedance is increased. The increase of the resistance has been compensated by increasing the thickness of the copper layer from 50 to 75 μm [3].

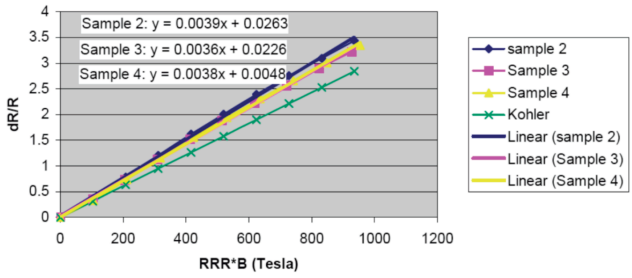

Figure 7: Electrical measurements of beam screen wall samples in magnetic fields compared to Kohler's formula. Courtesy of C. Rathjen [13].

## ANOMALOUS SKIN EFFECT (ASE)

The ASE theory attributes the anomalous increase of surface resistance of metals at low temperatures and high frequencies to the long mean free path of the conduction electrons. When the skin depth becomes much smaller than the mean free path, only a fraction of the conduction electrons moving almost parallel to the metal surface is effective in carrying the current and the classical theory breaks down. Some measurements were performed in Ref. [14], which were in relatively good agreement with predictions.

In the Normal Skin Effect (NSE), the skin depth and surface resistance are given respectively by

$$\delta = \sqrt{\frac{2\rho}{\omega \mu_0}} \quad \text{and} \quad R_s^{NSE} = \frac{\rho}{\delta} = \sqrt{\frac{\omega \mu_0 \rho}{2}}. \quad (18)$$

In the ASE, and approximate formula for the surface resistance was used in the past [15], which is valid when $\alpha \geq 3$ and which is given by

$$R_s^{ASE} = R_\infty \left(1 + 1.157\, \alpha^{-0.276}\right), \quad (19)$$

with

$$\alpha = \frac{3}{2}\left(\frac{\lambda}{\delta}\right)^2 = \frac{3\,\omega\,\mu_0}{4\,\rho^3}(\rho\,\lambda)^2, \quad (20)$$

$$R_\infty = \left[\frac{\sqrt{3}}{16\,\pi}\rho\,\lambda\,(\omega\,\mu_0)^2\right]^{1/3}$$

$$= 1.123 \times 10^{-3}\,\Omega\left(\frac{f}{\text{GHz}}\right)^{2/3}. \quad (21)$$

The parameter $R_\infty$ is independent of temperature and impurity, and $\rho\,\lambda = m\,v/(e^2\,N)$ is a characteristic of the metal, equal to $6.6\,10^{-16}\,\Omega\text{m}^2$ for copper. The relative increase of the heating power (assuming that the ASE formula is valid over the full frequency range) is given by (with $\sigma_z$ the rms bunch length in meters)

$$\frac{P_{ASE}}{P_{NSE}} = \frac{\int_{\omega=0}^{\omega=+\infty} d\omega\, R_s^{ASE}(\omega)\, e^{-\left(\frac{\omega\sigma_z}{c}\right)^2}}{\int_{\omega=0}^{\omega=+\infty} d\omega\, R_s^{NSE}(\omega)\, e^{-\left(\frac{\omega\sigma_z}{c}\right)^2}}. \quad (22)$$

Considering an rms bunch length $\sigma_z = 7.5$ cm leads to an increase by ~ 46% at injection (using the resistivity $1.8\,10^{-10}$ Ωm), an increase by ~ 11% at 8.33 T (using the resistivity $5.5\,10^{-10}$ Ωm), and an increase by ~ 4% at 20 T (using the resistivity $11.2\,10^{-10}$ Ωm). The plots of the surface impedance in both the NSE and ASE are shown in Fig. 8.

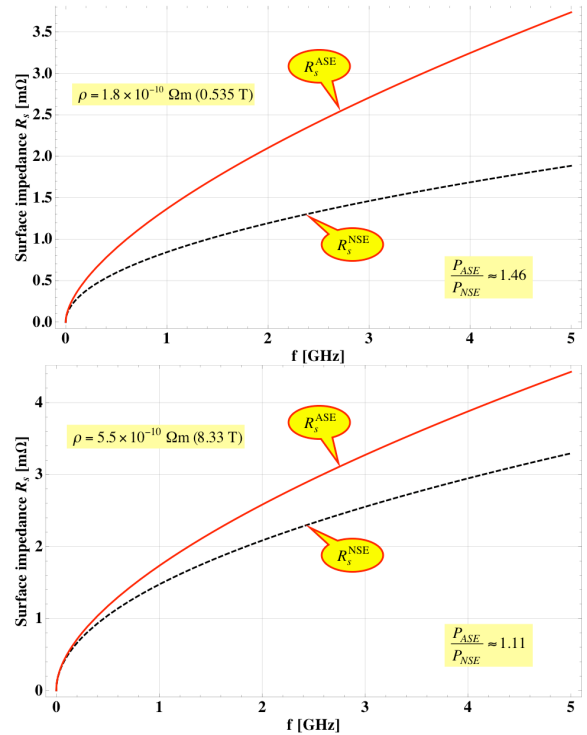

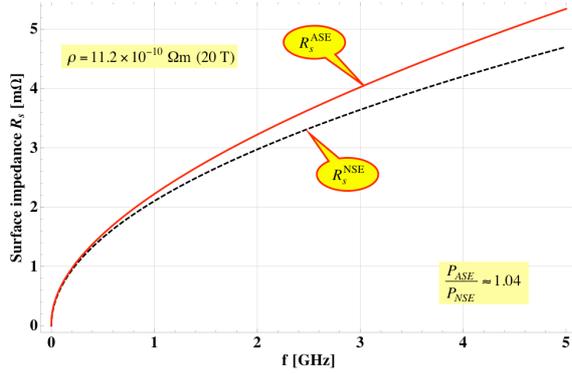

Figure 8: Surface impedance vs. frequency for both the NSE and ASE, assuming that they are valid over the full frequency range, for 0.535 T, 8.33 T and 20 T.

Sergio Calatroni implemented the exact (full) formula from Ref. [16], for the specular reflection of the electrons (which is the usual approximation; the diffuse contribution is close to this result). It is compared to the approximate formula of Eq. (19) in Fig. 9, where it can be seen that the exact formula converges to the NSE result at high temperature and to the limit $R_\infty$ at low temperature, whereas the approximate formula does not. Another interesting plot is shown in Fig. 10, where both the NSE and exact ASE formulae are plotted vs. *RRR* and magnetic field, for the particular frequency of 1 GHz. It is shown that for sufficiently high magnetic fields the result from NSE and ASE converge, and that in this case only the magneto-resistance needs to be taken into account as the ASE is small.

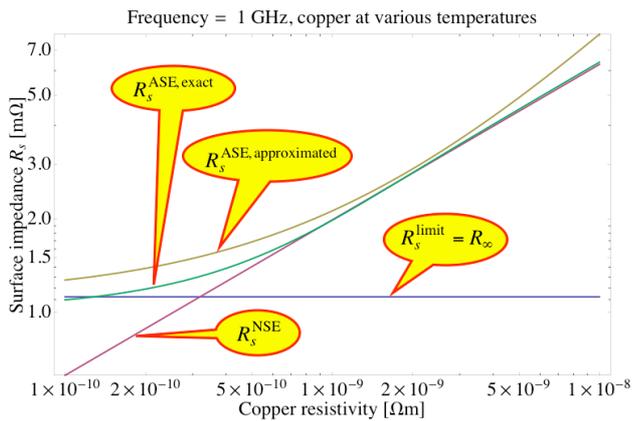

Figure 9: Surface impedance vs. copper resistivity (i.e. for different temperatures) and for the frequency of 1 GHz. The exact formula [16] is compared to the NSE and approximate ASE ones. This plot was made using the available *Mathematica* Notebook of Sergio Calatroni.

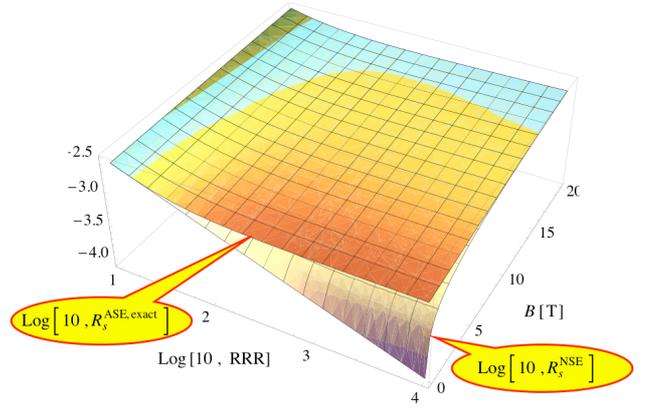

Figure 10: NSE and exact ASE formulae vs. *RRR* and magnetic field for the particular frequency of 1 GHz. This plot was made using the available *Mathematica* Notebook of Sergio Calatroni.

## CONCLUSIONS AND OUTLOOK

The magneto-resistance is the dominant effect for the beam screen in the HE-LHC with 20 T dipole magnets and the anomalous skin effect can be neglected. The beam screen copper resistivity at top energy increases from $\sim 5.5 \cdot 10^{-10}$ Ωm (at 7 TeV/c) to $\sim 11.2 \cdot 10^{-10}$ Ωm (at 16.5 TeV/c), i.e. by a factor $\sim 2$.

As the longitudinal and transverse impedances scale with the square root of the resistivity, they are larger by $\sim 40\%$, which should not be an issue for beam stability.

The total present power loss (from ohmic losses, pumping slots and the weld) is $\sim 150$ mW/m for one beam at 7 TeV/c. At 16.5 TeV/c, it increases only slightly to $\sim 175$ mW/m. Here again, no problem is expected.

In conclusion, no issues are anticipated for the beam screen if the beam energy is increased from 7 TeV/c to $\sim 16.5$ TeV/c. However, other impedance issues might arise with the collimators, whose gaps will be smaller and the transverse mode-coupling instability might be critical. It is worth reminding that at 7 TeV/c, the intensity threshold from the (single-bunch) transverse mode-coupling instability is estimated at (only) $\sim 2$ times the ultimate intensity. Furthermore, the threshold should even be smaller when taking into account the coupled-bunch effects [17], which are currently under study.

## ACKNOWLEDGEMENTS

It is a pleasure for me to thank S. Calatroni for all the very interesting discussions we had on magneto-resistance and anomalous skin effects. I would like also to thank V. Baglin, F. Caspers and N. Kos for helpful information.

## REFERENCES

[1] F. Ruggiero, Single-Beam Collective Effects in the LHC, Particle Accelerators, Vol. 50, pp. 83-104, 1995.

[2] N. Mounet, private communication, 2010.

[3] O. Brüning et al. (Editorial Board), LHC Design Report, Vol. I The LHC Main Ring, Chap. 5, CERN-2004-003, 4 June 2004.
[4] A. Mostacci, Beam-Wall Interaction in the LHC Liner, CERN-THESIS-2001-014, 2001.
[5] A. Mostacci, Beam-Wall Interaction in the LHC Liner: a Former PhD Student Experience, Francesco Ruggiero Memorial Symposium, CERN, 3 October 2007.
[6] W. Hofle, private communication, 2009.
[7] A.W. Chao, Physics of Collective Beam Instabilities in High Energy Accelerators, New York: Wiley, 371 p, 1993.
[8] F. Caspers et al., Surface Resistance Measurements and Estimate of the Beam-Induced Resistive Wall Heating of the LHC Dipole Beam Screen, LHC Project Report 307, 1999.
[9] A.W. Chao and M. Tigner (Editors), Handbook of Accelerator Physics and Engineering, 2$^{nd}$ Printing, p. 368.
[10] N. Luo and G.H. Miley, Kohler's Rule and Relaxation Rates in High-Tc Superconductors, Physica C 371 (2002) 259-269.
[11] L. Vos, Beam Vacuum Chamber Effects in the CERN Large Hadron Collider, CERN SPS/85-14 (DI-MST), 1985.
[12] Jeff Fitzgerald, Magnetoresistance, Physics 211A.
[13] A. Gerardin, Electrical Measurements of Beam Screen Wall Samples in Magnetic Fields, EST/SM-ME investigation report, EDMS N.329882.
[14] F. Caspers et al., Surface Resistance Measurements of LHC Dipole Beam Screen Samples, Proc. 7$^{th}$ EPAC, Vienna, Austria, 26-30 June 2000.
[15] W. Chou and F. Ruggiero, Anomalous Skin Effect and Resistive Wall Heating, LHC Project Note 2 (SL/AP), 1995.
[16] G.E.H. Reuter and E.H. Sondheimer, The Theory of the Anomalous Skin Effect in Metals, Proc. Royal Society (London), A195, 336 (1948).
[17] J.S. Berg and R.D. Ruth, Transverse Instabilities for Multiple Nonrigid Bunches in a Storage Ring, Physical Review E, Vol.52, N.3, September 1995.